# Rotational dynamics of optically trapped polymeric nanofibers

Antonio Alvaro Ranha Neves,[1,*] Andrea Camposeo,[1] Stefano Pagliara,[2] Rosalba Saija,[3] Ferdinando Borghese,[3] Paolo Denti,[3] Maria Antonia Iatì,[4] Roberto Cingolani,[1] Onofrio M. Maragò,[4,†] and Dario Pisignano[1,2]

[1] *National Nanotechnology Laboratory of CNR-INFM, IIT Research Unit, Università del Salento, via Arnesano, 73100, Lecce, Italy.*

[2] *Scuola Superiore ISUFI, Università del Salento, via Arnesano, 73100, Lecce, Italy.*

[3] *Dipartimento di Fisica della Materia e Ingegneria Elettronica, Università di Messina, Salita Sperone 31, 98166 Messina, Italy.*

[4] *Istituto per i Processi Chimico-Fisici (sez. Messina), CNR, Salita Sperone, 98158 Faro Superiore, Messina, Italy.*

**Abstract**

The optical trapping of polymeric nanofibers and the characterization of the rotational dynamics are reported. A strategy to apply a torque to a polymer nanofiber, by tilting the trapped fibers using a symmetrical linear polarized Gaussian beam is demonstrated. Rotation frequencies up to 10 Hz are measured, depending on the trapping power, the fiber length and the tilt angle. A comparison of the experimental rotation frequencies in the different trapping configurations with calculations based on optical trapping and rotation of linear nanostructures through a T-Matrix formalism, accurately reproduce the measured data, providing a comprehensive description of the trapping and rotation dynamics.

Optical forces are currently employed to study a range of chemical, physical and biological problems, by trapping microscale objects and measuring sub pico-Newton forces [1,2]. In particular, optical trapping of elongated nanoparticles, including nanowires [3] and nanotubes [4], is gaining an increasing interest because of the high shape anisotropy and unique physical properties of these systems. Among linear nanostructures, polymeric nanofibers are novel nanomaterials with many strategic applications ranging from scaffolding for tissue-engineering to integrated photonics [5,6] and electronics [7,8]. However, optical trapping and manipulation of polymer nanofibers has never been reported, despite the understanding of the optical forces and torques acting on these objects as well as their trapping dynamics might open a new range of applications, exploiting the polymeric fibers as local probes or active elements in microrheology [9] and microfluidics [10], and in next generation Photonic Force Microscopy [3]. Furthermore, the nanofibers are characterized by subwavelength diameters and lengths in the range 10-100 µm, therefore constituting ideal systems for studying effects occurring in the intermediate regime between the Rayleigh scattering and geometrical optics.

A laser beam can carry intrinsic (spin) or extrinsic (orbital) angular momentum, associated to the polarization and to the light beam phase structure, respectively [11]. Either trapping beams with elliptical polarization or with a rotating linear polarization can be exploited to apply a torque to trapped objects. Rotation in trapped particles can also be induced by exploiting the phase structure or the astigmatism of the trapping laser beam [12 and references therein]. The rotatable object can be spherical, exhibiting a birefringence or a slight absorption, or it can have more complex shapes, as in microfabricated propellers by two-photon polymerization [13] or cylinders with inclined faces [14].

In this work we trap polymeric nanofibers and characterize their rotational dynamics in different trapping configurations by a different method. We employ a strategy to rotate a dielectric cylinder with flat end faces, based on a non-rotating linear polarized Gaussian (TEM$_{00}$) beam, carrying neither intrinsic nor extrinsic angular momentum. The relevant advantages of this approach is that the trapped object does not need to be slightly absorptive, birefrigent, or specifically microfabricated. In addition, one does not need to manipulate the beam profile or polarization, being in this way more effective. This enables a detailed analysis of the torque acting on fibers, whose experimental results are compared with calculations of optical trapping and rotation of linear nanostructures through a full electromagnetic theory.

A detailed treatment of our modelling procedure is given in Ref.s [15,16]. Here we give a short account of calculations of the optical force and torque applied to polymer nanofibers (see Supplemental files for more details) [17]. Our starting point is the calculation of the field configuration in the focal region of a high numerical aperture (NA) objective lens in absence of particles [15]. Once the field is known, the radiation force and torque exerted on a trapped particle is calculated by resorting to momentum conservation for the combined system of field and particle. Then we use an approach based on the Transition (T-)Matrix formalism [18], to calculate the optical force and torque applied to the trapped object, of arbitrary size and symmetry, which is being modelled as an aggregate of spheres with size below the radiation wavelength. In the specific case of polymer nanofibers we calculate the radiation force ($\mathbf{F}_{rad}$) and torque ($\mathbf{M}_{rad}$) exerted by the optical tweezers, by modelling the nanostructures as linear chains of spheres with diameter, $D$, and length, $L$, equal to the fiber diameter and length, respectively. In particular, the calculations of the torque can be obtained for any orientation of the polymer fiber and for different trapping positions.

Dealing with quantitative comparisons between theory and experiments, a crucial issue to be addressed is the hydrodynamics of the trapped particle. For linear nanostructures (rigid rod-like structures), the viscous drag is described by an anisotropic hydrodynamic mobility tensor, whose components depend on the length of the linear structure ($L$) and on the length-to-diameter ratio, $p = L/D$ [19]. Symmetry considerations reduce the relevant hydrodynamics parameters to the translational, $\Gamma_\perp$ and $\Gamma_\parallel$, and rotational, $\Gamma_{Rot}$, mobilities [4,17]. Specifically when center-of-mass rotation is considered, the rotational mobility is:

$$\Gamma_{Rot} = \frac{3(\ln p + \delta_{Rot})}{\pi \eta L^3} \qquad (1)$$

where $\eta$ is the dynamical viscosity of the surrounding medium, and $\delta_{Rot}$ represents end corrections, calculated as polynomial of $(\ln 2p)^{-1}$ [19]. On the other hand, when the pivot point of the rotation is shifted by a value $\xi$ from the center-of-mass, we need to change $L$ with $L+2\xi$ in the rotational mobility [Eq. (1)]. For a rotating optically trapped polymer nanofiber, the radiation torque $\mathbf{M}_{rad}$ is counterbalanced by the hydrodynamic viscous torque (for the low Reynolds number regime) $\mathbf{M}_{hydro}=-\Omega/\Gamma_{Rot}\hat{\mathbf{n}}$ ($\hat{\mathbf{n}}$ is the rotation axis) and the fiber rotates at a constant rotation frequency:

$$\Omega = |\mathbf{M}_{rad}|\Gamma_{Rot}(L+2\xi) \qquad (2)$$

that is dependent both on the length of the fiber and on the pivot point position. Thus calculating the torque from our electromagnetic theory and using Eq. (1) and Eq. (2) yields the theoretical rotation frequency for the trapped fiber, that we can directly compare to measured experimental values.

Experimentally, our optical trap is custom-built on an inverted microscope (Zeiss Axiovert 40) as shown in Fig. 1(a), and based on a Ti:Sapphire laser ($\lambda$ =800 nm, Coherent). This is strongly focused to a diffraction-limited spot on the objective focal plane, by overfilling the back aperture of an oil-immersion infinity-corrected objective lens (100×/1.3, Zeiss Plan-Neofluar) [20]. Bright field images and videos are recorded by a charge coupled device camera using the same objective lens as the trapping laser. The polymeric nanofibers are fabricated by electrostatic spinning [21,22], exploiting a high electrostatic field (~0.9 kV cm$^{-1}$) to stretch a jet of polymer solution. Our samples are made by spinning a formic acid solution of poly(methylmethacrylate) (PMMA) [17] and dispersing the fibers (having diameters in the range 200-600 nm) in distilled water after complete solvent evaporation. The sample cell comprises a poly(dimethylsiloxane) (PDMS) chamber in conformal contact with a glass cover slip, thus defining a 100 μL volume of the water suspension of fibers. The cover slip is mounted on a piezoelectric stage, allowing travelling over 300 μm along each axis with nanometric spatial resolution.

The dynamics and tracking of the polymer fibers is investigated by means of the back-scattered light from the same laser used to trap the sample. In particular, the applied torque is related to the rotation of the trapped nanofiber, characterized by measuring the time evolution of the back-scattered light. Collecting this light by imaging the back focal plane of the microscope objective onto a silicon quadrant detector provides a direct, non-contact method to measure the drag torque. Upon Fourier processing, the particle rotation frequency is recovered by the power spectrum density [23] with high accuracy, and with larger bandwidth and better resolution than frame-by-frame video tracking.

Figures 1(b)-(e) highlight the stages of fiber trapping and rotation. The particle is picked up from glass with low optical power [< 50 mW, Fig. 1(b)], and taken to a distance from the cover slip slightly larger than the fiber length. Once in this position the fiber stands upright, aligning its longitudinal axis to that of the optical axis of the laser beam [Fig. 1(c)]. We then increase the trapping power (100-400 mW) to confine the fiber in a stiff trap. In this configuration we can translate the beam and the fiber along all the three axes. By approaching the cover slip to the fiber using the piezo-stage, the glass surface is led again in contact with the bottom tip of the fiber, which starts to tilt by an angle $\theta$ [Fig. 1(d)]. Above a critical angle depending on the trapping power, the fiber begins to rotate at constant rate ($\Omega$). For some samples, we can also finally lower the piezo-stage, thus leaving the fiber not in contact with the glass surface while continuing to rotate in its tilted configuration [Fig. 1(e)].

In order to characterize the trapped fibers, we first test the alignment between the nanofiber and the polarization direction of the trapping beam for the configuration depicted in Fig. 1(d). To this aim we tilt the trapped cylindrical object near the surface, using low optical power and thus not causing any rotation, and we use a $\lambda/2$ plate to rotate the trapping beam polarization. The asymmetric backscattered light off the nanofiber is detected on the quadrant photodiode for each $\lambda/2$ plate angle, $\psi$. Upon decomposing the signal in top-bottom and left-right pairs, we can determine the orientation angle of the new equilibrium position with respect to $\psi$. We find that the fiber aligns with the local polarization axis of the trapping field (Fig. 2), in agreement with previous results by Bishop *et al.* using glass rods [24] and recently reported analytical calculations of spheroids [25]. Instead, with the trapped object pulled away from the cover slip, as schematized in Fig. 1(c), the fiber aligns along the optical axis, and no orientation with respect to the incident polarization or rotation is observed within the experimental error. Such lack of rotation for the fiber when aligned to the optical axis is an indication of the isotropy of the cylinder and end faces, as observed in the scanning electron microscopy (SEM) micrograph in the inset of Fig. 3(a) [13 and references therein].

We then analyze the dynamics of tilted fibers by measuring the rotating frequency as a function of the incident optical power. A time series of the quadrant photodetector signal over 20 s is used to determine the rotation frequency, from the frequency peak of the power spectrum [Fig. 3(b)]. Since only one sharp and symmetrical peak is detected along with its harmonics in the power spectral density, we conclude that the observed frequency is that of a continuous rotation of the fiber without nutation. We find that the rotation frequency increases linearly with the trapping power [linear fits through the origin shown in Fig. 3(c)], which rules out the occurrence of nonlinear effects in the investigated experimental range. Moreover, $\Omega$ decreases upon increasing the fiber length, as expected from the dependence of the rotation frequency on $L$ [Eq. (2)] [17].

Finally, we investigate the dependence of the rotation frequency on the tilting angle ($\theta$). To this aim, we keep the rotating fiber at a constant power (115 mW) and vary the fiber-to-cover slip distance using the piezo-stage at sub-micron intervals. The angle $\theta$ is calculated from the fiber length and the fiber-to-cover slip distance. In Fig. 4, we display the rotation frequency vs. $\sin\theta$. The rotation frequency is almost constant for angles up to 82° ($\sin\theta$=0.99) and the calculated frequencies, based on a model of a rotating fiber around its end-point ($\xi \approx L/2$) ([17]), well describe the measured values. An increase of the rotation frequency is observed for angles approaching $\sin\theta$=1, an effect that is not related to the increasing tilt, that would cause a decrease of the rotating frequency if the pivot point is unchanged (Fig. 4). We attribute the increase of the rotation frequency for high tilt angles to a progressive shift of the trapping point from the fiber tip towards the fiber center-of-mass. In fact, our calculation reproduce quite well the experimental results when assuming a progressive shift of the trapping point from the fiber tip towards its center-of-mass mass (up to $\xi \approx L/4$).

In conclusion we demonstrated optical trapping and manipulation of polymeric nanofibers, introducing the control of rotation over elongated nanostructures by tilting the trapped fiber, which allows to achieve rotation frequencies up to 10 Hz. The measured rotation frequencies in the different trapping configurations are well reproduced by calculations based on a T-Matrix formalism for optical force and torque. The manipulation of this novel class of nanomaterials hold promises for a wealth of applications, such as photonic circuits or microfluidic devices, that can benefit from the controlled manipulation and rotation of the nanofibers, and the assembly of active polymeric fibers in ordered arrays. In particular the control over length and size makes polymer nanofibers ideal probes in next generation Photonic Force Microscopy.

Acknowledgments

This work was partially supported by the Italian Minister of University and Research through the FIRB programs RBIN045NMB and RBIP06SH3W, and by the Apulia Regional Strategic Project PS_144. The authors gratefully acknowledge R. Stabile for the SEM images.

**List of Figures**

Fig. 1. (color online) (a) Scheme of the optical tweezers set-up with detection using backscattered light. The arrows indicate the light paths. (b)-(e): Optical rotation of the polymer fiber, schematized (top), and imaged (bottom). The fiber main axis is tilted by $\theta$ from the optical axis, the trapping point is shifted by $\xi$ from the center-of-mass (CM). Scale bar 2.5 µm

Fig. 2. (color online) Non-rotating tilted fiber aligning with external linear polarization as a function of waveplate angle, $\psi$. Continuous line (blue) represents the x-axis position (left vertical scale), and the dashed line (red) the y-axis position (right vertical scale), projected onto the quadrant photodiode. Inset: Micrographs of the tilted nanofiber for $\psi$ = 120° (a), 140° (b), and 160° (c). Scale bar = 5 µm.

Fig. 3. (color online) (a) SEM of typical polymer fibers used in the rotation experiments, scale bar = 2 µm. (b) Typical peak of power spectral density corresponding to the rotating frequency of the trapped fiber ($L$ =8.3 µm) at optical power 113 mW. (c) Trapped rotating fiber frequency vs. trapping power, for nanofibers of different lengths, with their respective fits through the origin. Power values measured for light before entering the microscope.

Fig. 4. (color online) Comparison of the measured (full blue squares) and calculated (empty red circles and empty red diamonds) trapping frequency vs. sine of the tilt angle, $\theta$, for a fiber with length $L$ = 8.8 µm. The rotation frequencies are calculated by assuming a trapping point close to the fiber tip (empty red circles) or a progressive shift towards the fiber center of mass (empty red diamond). The shift between two consecutive calculated points is of 0.45 µm.

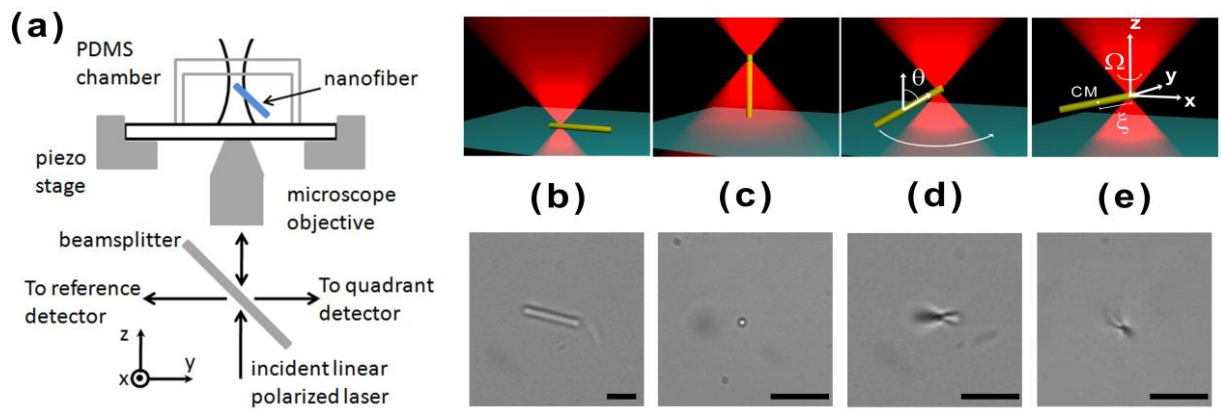

Fig. 1. (color online) (a) Scheme of the optical tweezers set-up with detection using backscattered light. The arrows indicate the light paths. (b)-(e): Optical rotation of the polymer fiber, schematized (top), and imaged (bottom). The fiber main axis is tilted by $\theta$ from the optical axis, the trapping point is shifted by $\xi$ from the center-of-mass (CM). Scale bar 2.5 µm

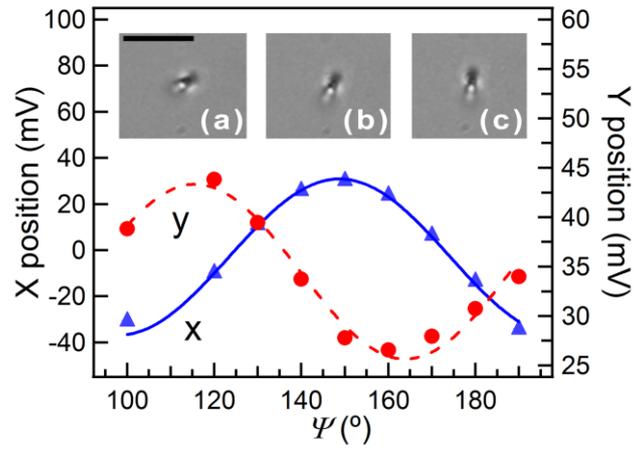

Fig. 2. (color online) Non-rotating tilted fiber aligning with external linear polarization as a function of waveplate angle, $\psi$. Continuous line (blue) represents the x-axis position (left vertical scale), and the dashed line (red) the y-axis position (right vertical scale), projected onto the quadrant photodiode. Inset: Micrographs of the tilted nanofiber for $\psi$ = 120° (a), 140° (b), and 160° (c). Scale bar = 5 µm.

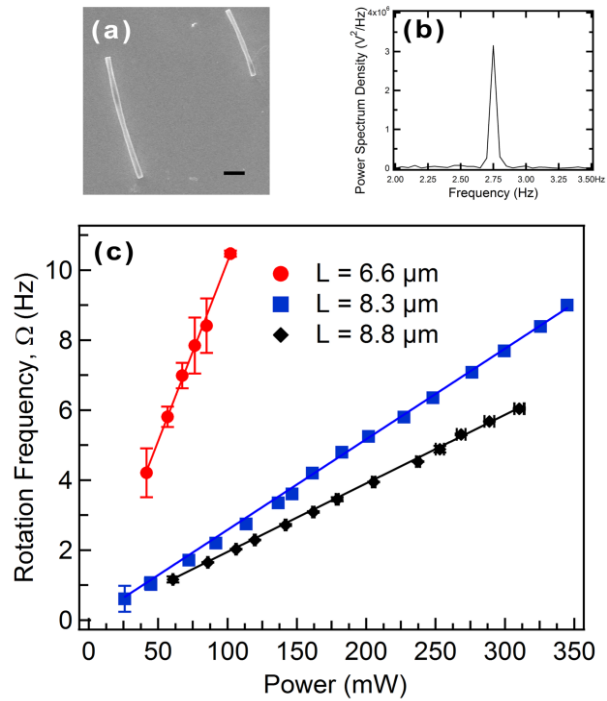

Fig. 3. (color online) (a) SEM of typical polymer fibers used in the rotation experiments, scale bar = 2 µm. (b) Typical peak of power spectral density corresponding to the rotating frequency of the trapped fiber ($L$ =8.3 µm) at optical power 113 mW. (c) Trapped rotating fiber frequency vs. trapping power, for nanofibers of different lengths, with their respective fits through the origin. Power values measured for light before entering the microscope.

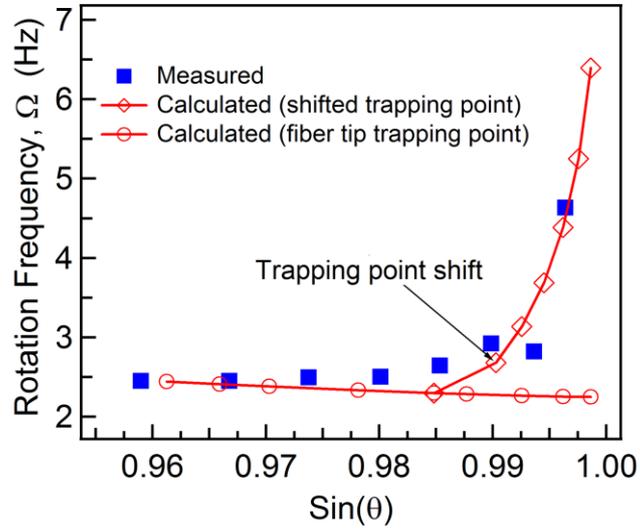

Fig. 4. (color online) Comparison of the measured (full blue squares) and calculated (empty red circles and empty red diamonds) trapping frequency vs. sine of the tilt angle, $\theta$, for a fiber with length $L$ = 8.8 µm. The rotation frequencies are calculated by assuming a trapping point close to the fiber tip (empty red circles) or a progressive shift towards the fiber center of mass (empty red diamond). The shift between two consecutive calculated points is of 0.45 µm.

# Rotational dynamics of optically trapped polymeric nanofibers (EPAPS)

Antonio Alvaro Ranha Neves,[1,*] Andrea Camposeo,[1] Stefano Pagliara,[2] Rosalba Saija,[3] Ferdinando Borghese,[3] Paolo Denti,[3] Maria Antonia Iatì,[4] Roberto Cingolani,[1] Onofrio M. Maragò,[4,†] and Dario Pisignano[1,2]

[1] *National Nanotechnology Laboratory of CNR-INFM, IIT Research Unit, Università del Salento, via Arnesano, 73100, Lecce, Italy.*

[2] *Scuola Superiore ISUFI, Università del Salento, via Arnesano, 73100, Lecce, Italy.*

[3] *Dipartimento di Fisica della Materia e Ingegneria Elettronica, Università di Messina, Salita Sperone 31, 98166 Messina, Italy.*

[4] *CNR-Istituto per i Processi Chimico-Fisici (Messina), Salita Sperone, 98158 Faro Superiore, Messina, Italy.*

* Corresponding author: antonio.neves@unisalento.it

† marago@its.me.cnr.it

**1. Radiation Force and Torque**

Light forces are generated by the scattering of electromagnetic fields incident on a particle, hence the quantitative understanding of optical trapping has to rely on the scattering theory of electromagnetic radiation [1]. The difficulties arising from the use of the full scattering theory are generally overcome by solving the problem in different regimes depending on the size of the scatterer. Moreover the models traditionally used for calculating optical forces are based on approximations which often limit the discussion only to spherical particles. On the contrary, in order to calculate the radiation force [2] and torque [3,4] we use the full scattering theory in the framework of the transition matrix (T-matrix) approach. In fact, this approach is quite general as it applies to particles of any shape and refractive index for any choice of the wavelength. Our starting point is the calculation of the field configuration in the focal region of a high numerical aperture (NA) objective lens in absence of any particle, using the procedure originally formulated by Richards and Wolf [5]. The resulting field is considered as the field incident on the particles, and the radiation force and torque exerted on any particle within the region is calculated by resorting to momentum conservation for the combined system of field and particles. As a result the optical force and torque exerted on a particle turn out to be given by the integrals [2,3]:

$$\mathbf{F}_{\text{Rad}} = r^2 \int_{\Omega} \hat{\mathbf{r}} \cdot \langle \mathsf{T}_{\text{M}} \rangle \, d\Omega, \tag{1}$$

$$\mathbf{M}_{\text{Rad}} = -r^3 \int_{\Omega} \hat{\mathbf{r}} \cdot \langle \mathsf{T}_{\text{M}} \rangle \times \hat{\mathbf{r}} \, d\Omega, \tag{2}$$

where the integration is over the full solid angle, $r$ is the radius of a large (possibly infinite) sphere surrounding the particle centre, and $\langle \mathsf{T}_{\text{M}} \rangle$ is the time averaged Maxwell stress tensor:

$$\langle \mathsf{T}_{\text{M}} \rangle = \frac{1}{8\pi} \text{Re}[n^2 \mathbf{E} \otimes \mathbf{E}^* + \mathbf{B} \otimes \mathbf{B}^* - \frac{1}{2}(|\mathbf{E}|^2 + |\mathbf{B}|^2)\mathbf{I}], \tag{3}$$

where $\otimes$ denotes dyadic product, $\mathbf{I}$ is the unit dyadic and $n$ is the refractive index of the particle. When the incident field is a polarized plane wave, the components of the radiation force along the direction of the unit vector $\hat{\mathbf{v}}_\xi$ are given by [2]:

$$\mathbf{F}_{\mathrm{Rad}\,\xi} = -\frac{r^2}{16\pi}\,\mathrm{Re}\int(\hat{\mathbf{r}}\cdot\hat{\mathbf{v}}_\xi)[n^2(|\mathbf{E}_S|^2 + 2\mathbf{E}_I\cdot\mathbf{E}_S) + (|\mathbf{B}_S|^2 + 2\mathbf{B}_I\cdot\mathbf{B}_S)]\,d\Omega, \qquad (4)$$

where $\mathbf{E}_I$ and $\mathbf{B}_I$ are the incident fields, while $\mathbf{E}_S$ and $\mathbf{B}_S$ are the fields scattered by the particle. In turn the radiation torque takes on the form:

$$\mathbf{M}_{\mathrm{Rad}} = \frac{1}{8\pi}r^3\,\mathrm{Re}\int n^2\hat{\mathbf{r}}\cdot(\mathbf{E}_I+\mathbf{E}_S)(\mathbf{E}_I+\mathbf{E}_S)\times\hat{\mathbf{r}}\,d\Omega. \qquad (5)$$

Expanding the incident field in a series of vector spherical harmonics with (known) amplitudes $W^p_{I\,lm}$, the scattered field can be expanded on the same basis with amplitudes $A^{p'}_{l'm'}$. The relation between the two amplitudes is given by $A^{p'}_{l'm'} = \sum_{plm} S^{p'p}_{l'm'lm} W^p_{I\,lm}$, where $S^{p'p}_{l'm'lm}$ is the T-matrix of the particle. In this framework, several kinds of non-spherical particles can be modeled as aggregates of spherical scatterers with size below the radiation wavelength. The elements of the T-matrix are calculated in a given frame of reference through the inversion of the matrix of the linear system obtained by imposing to the fields boundary conditions across each spherical surface [1]. Here we stress that each element of the T-matrix turns out to be independent both on the direction of propagation and on the polarization of the incident field. Thus they do not change when the incident field is a superposition of plane waves with different direction of propagation, i.e. for the description of a focused laser beam in the angular spectrum representation [6]. In our calculations, the polymer nanofibers are modeled as chains of spheres, with diameter corresponding to the fiber diameter. The chain length corresponds to the fiber one. In Fig. 1(a)-(b) a schematic representation of the geometrical configuration of the modeled system is reported. The results of typical calculations of the optical torque exerted on a tilted polymer fiber with $L=9$ μm and $D=0.45$ μm is displayed in Fig. 1(c). The torque values for the case of a trap centered with the fiber center-of-mass (open squares) are much higher than values for a fiber trapped and rotating around an end-point (open circles).

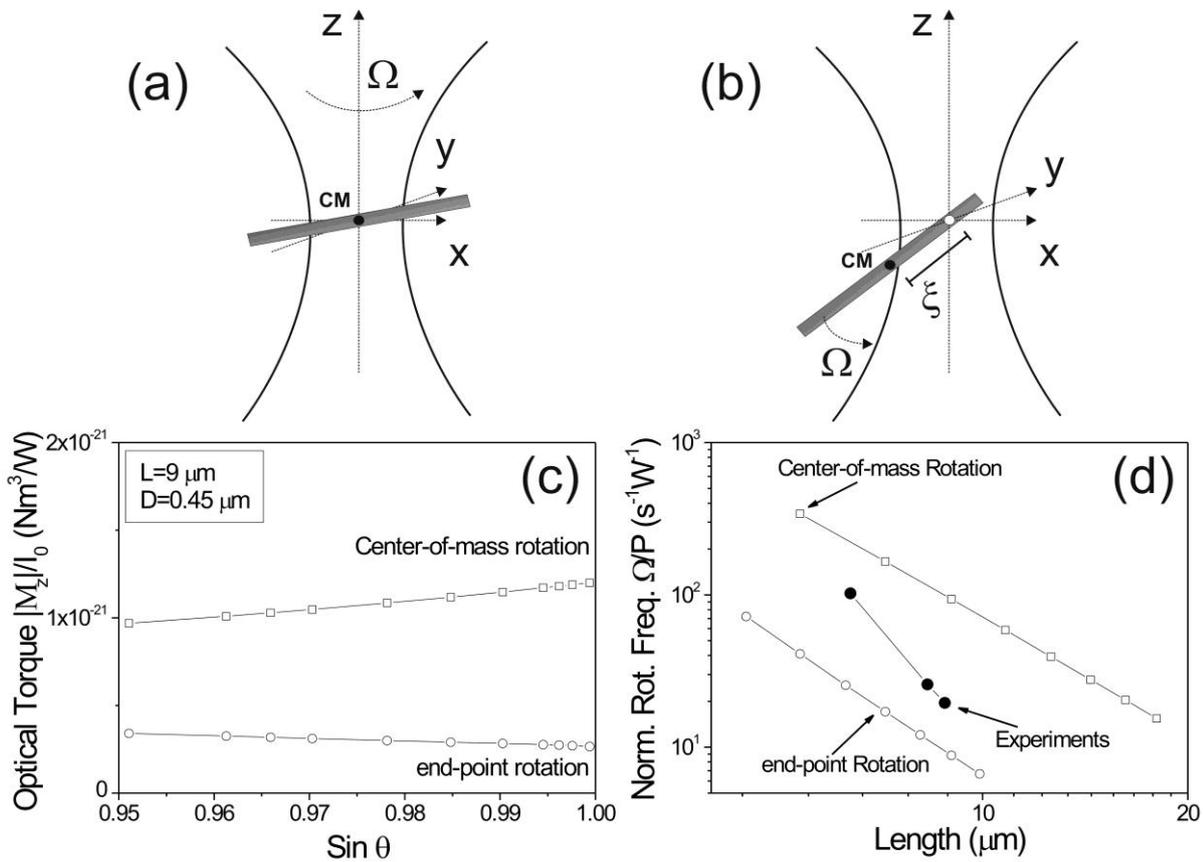

Fig. 1. (a) Sketch of the geometric configuration for a fiber trapped and rotated about its center-of-mass (black dot). Rotation occurs in the xy plane. (b) Geometry of the optical trapping of a fiber rotated about a point shifted by $\xi$ towards the edge of the fiber (white dot). The rotational mobility is controlled by the parameter $L+2\xi$ (see text). (c) Dependence with tilting angle $\theta$ of the calculated optical axis (z-)component of the optical torque (normalized to the intensity of radiation) exerted on a polymer fiber with $L$=9 μm and $D$=0.45 μm modeled as a linear chain of spheres. Torque values for a trap centered on the fiber center-of-mass (open squares) are much higher than values for a fiber rotating around an end-point (open circles). (d) Calculated rotation frequencies (normalized to incident power) of tilted polymer fibers as a function of their length $L$ and fixed diameter $D$=0.45 μm for the two configurations of center-of-mass rotation (open squares) and end-point rotation (open circles). Tilting was fixed at $\theta$=85° from the optical axis. The experimental points (filled circles) are between the two extreme cases, center-of-mass and end-point trapping.

**2. Hydrodynamics**

When dealing with quantitative measurements of optical trapping and rotation on polymer nanofibers, a crucial issue to be considered is the hydrodynamics of the trapped particle. For rigid rod-like structures of length $L$ and diameter $D$, the viscous drag is described by an anisotropic hydrodynamic mobility tensor [7,8], the components of which depend on the length of the structure ($L$) and on the length-to-diameter ratio $p = L/D$ as [7]:

$$\Gamma_{\perp} = \frac{\ln p + \delta_{\perp}}{4\pi\eta L}, \; \Gamma_{\parallel} = \frac{\ln p + \delta_{\parallel}}{2\pi\eta L}, \; \Gamma_{Rot} = \frac{3(\ln p + \delta_{Rot})}{\pi\eta L^3}, \tag{6}$$

where $\Gamma_{\perp}$ and $\Gamma_{\parallel}$ are the translational mobilities, transverse and parallel to the main axis respectively, $\Gamma_{Rot}$ is the rotational mobility about the center-of-mass, $\eta$ is the water dynamical viscosity, and $\delta_i$ are end corrections (calculated in [7] as polynomial of $(\ln 2p)^{-1}$). Note that when rotation of a linear nanostructure occurs about a point shifted by $\xi$ from the center-of-mass, the rotational mobility changes as:

$$\Gamma_{Rot}(L+2\xi) = \frac{3(\ln \tilde{p} + \tilde{\delta}_{Rot})}{\pi\eta(L+2\xi)^3}, \tag{7}$$

where $\tilde{p} = (L+2\xi)/D$ is an effective length-to-diameter ratio and $\tilde{\delta}_{Rot}$ is an effective end correction that takes into account the shift of the rotation pivot point with respect to the center-of-mass. In our experimental situation of a rotating trapped polymer fiber, an equilibrium is reached when the radiation torque $\mathbf{M}_{rad}$ is counterbalanced by the hydrodynamic viscous torque $\mathbf{M}_{hydro} = -\Omega/\Gamma_{Rot}\hat{\mathbf{n}}$ ($\hat{\mathbf{n}}$ is the rotation axis) and the fiber rotates at a constant rotation frequency $\Omega = |\mathbf{M}_{rad}|\Gamma_{Rot}(L+2\xi)$ that is dependent both on the length of the fiber and on the pivot point position. This relation holds when the fiber rotates in a plane orthogonal to the optical axis, i.e. for a situation fulfilled in our experiments where the tilting angle is close to 90° i.e. $\sin\theta \approx 1$. In Fig. 1(d) we show rotation frequencies (normalized to incident power) as a function of fiber length calculated for polymer fibers of different length $L$ and fixed diameter $D$ =0.45 μm, for the two situations of center-of-mass rotation (open squares) and end-point rotation (open circles). Tilting was fixed at $\theta$ =85° from the optical axis. The experimental points (filled circles) are between the two extreme cases showing that the trapping (and rotation) point is shifted from the end-point of the fiber. This shows that the rotation frequency ($\Omega$) depends dramatically on the trapping point.

## 3. Realization of Nanofibers

Electrostatic spinning (ES) is performed by processing a solution of poly(methylmethacrylate) (PMMA) dissolved in formic acid with relative concentration of 26% (w/w). Formic acid, because of its high conductivity and dielectric constant, allows the realization of uniform and beads-free fibers with a narrow dispersion diameters in the sub-micrometer range. In a typical electrospinning process a 0.5 ml of PMMA solution is loaded into a 1.0 mL plastic syringe tipped with a 19-gauge stainless steel needle. The positive lead from a high voltage supply (XRM30P, Gamma High Voltage Research Inc., Ormond Beach, FL) is connected to the metal needle applying a bias of 9 kV. The solution is injected at the end of the needle at a constant rate of 10 μL/min by a syringe pump (33 Dual Syringe Pump, Harvard Apparatus Inc., Holliston, MA), which prevents dripping at the end of the metallic capillary. Fibers are collected as non-woven mat on an aluminum collector negatively biased at -2 kV and placed at a distance of 12 cm from the needle. All the electrospinning experiments are performed at room temperature with air humidity about 40%. Finally, fibers are mechanically removed from the collector and stored in a vial containing distilled water. To allow the subsequent separation and fragmentation, the suspension of fibers in distilled water is sonicated for 1 hour at 25 °C before the trapping experiments.